\documentclass[conference, 9pt]{IEEEtran}
\IEEEoverridecommandlockouts 

\usepackage{multicol}
\usepackage{pifont}
\usepackage{placeins}
\usepackage{setspace}
\usepackage{soul}
\usepackage{threeparttable}
\usepackage{times}
\usepackage{url}
\usepackage{upgreek}
\usepackage{verbatim}
\usepackage{wrapfig}
\usepackage[usenames,dvipsnames]{xcolor}
\usepackage{tabularx}
\usepackage{supertabular}
\usepackage{tikz}
\usepackage{blindtext}
\usepackage{hhline} 
\usepackage[export]{adjustbox}

\definecolor{lightgreen}{RGB}{198, 224, 183}
\definecolor{lightorange}{RGB}{242, 178, 136}

\graphicspath{{./_fig/}}

\makeatletter
\renewcommand\footnoterule{%
  \kern-3\p@
  \hrule\@width0.4\columnwidth
  \kern2.6\p@}
  \makeatother

\usepackage[hang,flushmargin]{footmisc}
\usepackage{bbding}
\usepackage{lettrine}
\usepackage{xcolor,colortbl}

\definecolor{cyan}{RGB}{213,229,255}
\definecolor{orangeShallow}{RGB}{255,190,0}

\definecolor{new_blue}{HTML}{4b5cc4}

\usepackage{subfigure}
\usepackage{tikz}
\usepackage{xcolor}
\newcommand*\circled[1]{\tikz[baseline=(char.base)]{
            \node[shape=circle,fill,inner sep=1pt,scale=0.8] (char) {\textcolor{white}{#1}};}}
\usepackage{blkarray}                                      
\usepackage{algpseudocode}                                 
\usepackage[linesnumbered,ruled,vlined]{algorithm2e}
\usepackage{graphicx}                                      
\usepackage{amsmath}
\usepackage{amssymb}
\usepackage{amsfonts}
\usepackage{amsthm}
\usepackage[mathcal]{eucal}
\usepackage{mathrsfs}
\usepackage{booktabs}
\usepackage{enumerate}
\usepackage{multirow}
\usepackage{cite}                                          
\usepackage{comment}                                       
\usepackage{soul}                                          
\soulregister\cite7
\soulregister\ref7
\soulregister\pageref7
\usepackage{etoolbox}                                      
\usepackage{url}
\usepackage{nth}                                           
\usepackage{bm}                                            
\usepackage{courier}
\usepackage{balance}
\usepackage{threeparttable}
\usepackage{xcolor,colortbl}
\usepackage{footnote}

\usepackage{verbatim}
\usepackage[bookmarks=false]{hyperref}
\hypersetup{
    colorlinks = true,
    citecolor  = blue,
    linkcolor  = blue,
    urlcolor   = blue,
}
\usepackage{tikz}
\usetikzlibrary{patterns,snakes}
\usetikzlibrary{positioning,calc,fit,decorations.pathmorphing,shapes.geometric, shapes.gates.logic.US, calc}
\usetikzlibrary{arrows,arrows.meta,decorations.markings,shapes,shapes.arrows}
\usetikzlibrary{decorations,decorations.pathreplacing}
\usetikzlibrary{backgrounds}
\usepackage{filecontents}                                  
\usepackage{pgfplots}
\usepackage{pgfplotstable}
\usepackage{scalefnt}
\pgfplotsset{compat=newest}
\usepackage{caption}
\usepackage{cleveref}
\Crefformat{figure}{Fig.~#2#1#3}                           
\Crefname{subfigure}{Fig.}{Figs.}
\Crefname{figure}{Fig.}{Figs.}
\Crefformat{table}{TABLE~#2#1#3}                           
\usepackage[figuresright]{rotating}

\definecolor{CUHKorange}{RGB}{244,106,18} 
\definecolor{CUHKblue}{RGB}{0,111,190}    
\definecolor{CUHKgreen}{RGB}{0,127,128}   
\definecolor{CUHKred}{RGB}{228,46,36}     
\definecolor{CUHKyellow}{RGB}{198,148,34} 
\definecolor{CUHKdark}{RGB}{114,44,114}   
\definecolor{CUHKmiddle}{RGB}{144,44,144} 
\definecolor{CUHKlight}{RGB}{167,44,167}

\definecolor{USTgold}{RGB}{153,102,0}
\definecolor{USTyellow}{RGB}{204,153,0}
\definecolor{USTyellowlight}{RGB}{255,212,0}
\definecolor{USTorange}{RGB}{255,166,26}
\definecolor{USTpink}{RGB}{255,157,157}
\definecolor{USTblue}{RGB}{0,51,102}
\definecolor{USTmiddle}{RGB}{0,116,188}
\definecolor{USTlight}{RGB}{99,202,225}
\definecolor{USTgray}{RGB}{204,204,204}
\definecolor{USTred}{RGB}{237,27,47}
\definecolor{USTdarkred}{RGB}{124,35,72}

\usepackage{tcolorbox}
\tcbuselibrary{skins,breakable}
    {\endtcolorbox}

\crefname{mytheorem}{Theorem}{Theorems}
\crefname{mylemma}{Lemma}{Lemmas}
\crefname{myclaim}{Claim}{Claims}
\crefname{myproperty}{Property}{Properties}
\crefname{mycorollary}{Corollary}{Corollaries}

\algrenewcommand\textproc{\texttt}

\makeatletter
\let\OldStatex\Statex
\renewcommand{\Statex}[1][3]{%
  \setlength\@tempdima{\algorithmicindent}%
  \OldStatex\hskip\dimexpr#1\@tempdima\relax
}
\makeatother

\RequirePackage[normalem]{ulem} 
\RequirePackage{color}\definecolor{RED}{rgb}{1,0,0}\definecolor{BLUE}{rgb}{0,0,1} 

\usepackage{amsmath}
\usepackage{amsfonts}
\usepackage{url}
\usepackage{bm}
\usepackage{comment}
\usepackage{multirow}
\usepackage{color}
\usepackage{tabu}
\usepackage{mdframed}
\usepackage{syntax}
\usepackage{fancyvrb}
\usepackage{xcolor} 

\hypersetup{hidelinks}

\usepackage{multirow}
\usepackage{bbding}
\usepackage{mathtools}
\usepackage{algorithm2e}

\usepackage{paralist}
\usepackage{stmaryrd}
\usepackage{tikz}
\usepackage[referable]{threeparttablex}
\usepackage{tabularx}
\usepackage{booktabs}
\usepackage{array}
\usepackage{fancyhdr}
\usepackage{pifont}
\usepackage{balance}
\usepackage{xcolor}

\usepackage{tablefootnote}
\usepackage[normalem]{ulem}
\usepackage{graphicx}

\usepackage{tikz}
\usepackage{xcolor}

\usepackage{cleveref}

\usepackage{graphicx}

\usepackage{pdfpages}
\usepackage{changepage}

\definecolor{princetonorange}{RGB}{255,143,0}

\definecolor{lightgreen}{RGB}{198, 224, 183}

\definecolor{lightred}{RGB}{240, 205, 176}

\definecolor{newblue}{RGB}{66, 173, 245}
\definecolor{cyan}{RGB}{213,229,255}
\definecolor{yellow}{RGB}{253, 243, 208}
\definecolor{orangeShallow}{RGB}{255,190,0}
\definecolor{ShallowYellow}{RGB}{249,241,204}
\definecolor{ShallowGreen}{RGB}{222,237,214}
\definecolor{ShallowOrange}{RGB}{250,229,212}
\definecolor{NormalGreen}{RGB}{139, 195, 74}
\definecolor{ShallowPurple}{RGB}{232, 223, 243}

\usepackage{colortbl}

\begin{document}



\setcounter{page}{1}
\twocolumn

\title{HLSDebugger: Identification and Correction of Logic Bugs\\in HLS Code with LLM Solutions}

\author{
    \IEEEauthorblockN{
        Jing Wang,
        Shang Liu,
        Yao Lu,
        Zhiyao Xie\IEEEauthorrefmark{1}\thanks{ \IEEEauthorrefmark{1}Corresponding Author.}
    }
    \vspace{.01in}
    \IEEEauthorblockA{Hong Kong University of Science and Technology}
    \vspace{.01in}
    \IEEEauthorblockA{
        \{jwangjw, sliudx, yludf\}@connect.ust.hk,\quad
        eezhiyao@ust.hk
    }
}

\maketitle

\begin{abstract}

High-level synthesis (HLS) accelerates hardware design by enabling the automatic translation of high-level descriptions into efficient hardware implementations.
However, debugging HLS code is a challenging and labor-intensive task, especially for novice circuit designers or software engineers without sufficient hardware
domain knowledge. The recent emergence of Large Language Models (LLMs) is promising in automating the HLS debugging process. 
Despite the great potential, three key challenges persist when applying LLMs to HLS logic debugging:
1)  High-quality circuit data for training LLMs is scarce, posing a significant challenge. 2) Debugging logic bugs in hardware is inherently more complex than identifying software bugs with existing golden test cases. 
3) The absence of reliable test cases requires multi-tasking solutions, performing both bug identification and correction. 
In this work, we propose a customized solution named HLSDebugger\footnote{Open-sourced at
https://github.com/hkust-zhiyao/HLSDebugger}, to address the challenges. HLSDebugger first generates and releases a large labeled dataset with 300K data samples, targeting HLS logic bugs. The HLSDebugger model adopts an encoder-decoder structure, performing bug location identification, bug type prediction, and bug correction with the same model. HLSDebugger significantly outperforms advanced LLMs like GPT-4 in bug identification and by more than $3\times$ in bug correction. It makes a substantial advancement in the exploration of automated debugging of HLS code.


\end{abstract}



\IEEEpeerreviewmaketitle


\section{Introduction}\label{sec:intro}

High-Level Synthesis (HLS) has revolutionized the hardware design process by allowing designers to define hardware functionality using high-level programming languages, such as C++ or SystemC. Such a high-level abstraction of circuits accelerates the design process and thus enables rapid prototyping and agile development of hardware. However, as with any development process, debugging is a critical component, which is labor-intensive and requires knowledge in both software and hardware domains. In real-world situations, many pre-silicon logic bugs will be inadvertently introduced by hardware designers when crafting the HLS version of a design. Such logic bugs can elude HLS and static analysis tools (e.g., Infer) \cite{10473893}, causing unintended functionalities.



\begin{figure}
    \centering
    \subfigure[Traditional LLM-based solutions for debugging tasks generate bug identification and correction sequentially in the output text.]{
    \centering
    \includegraphics[width=0.48\textwidth]{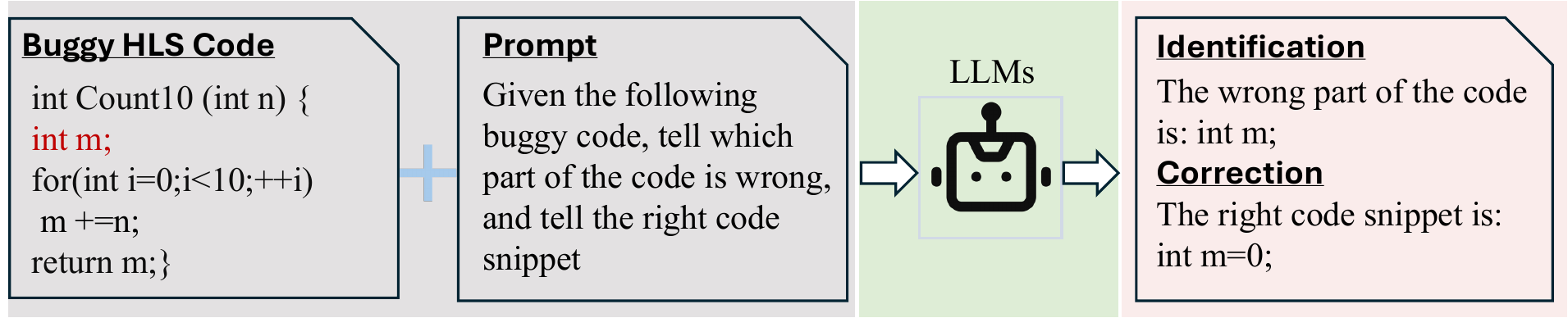}
    \label{fig:traditional-LLMs-Solution}
    }
    \subfigure[Our customized bug identification solution, HLSDebugger, directly generates bug location and provides bug corrections without requiring a carefully crafted prompt. The HLSDebugger can also utilize information (e.g, Bug Location Indication) from HLS tools to assist the bug correction process.]{
    \centering
    \includegraphics[width=0.48\textwidth]{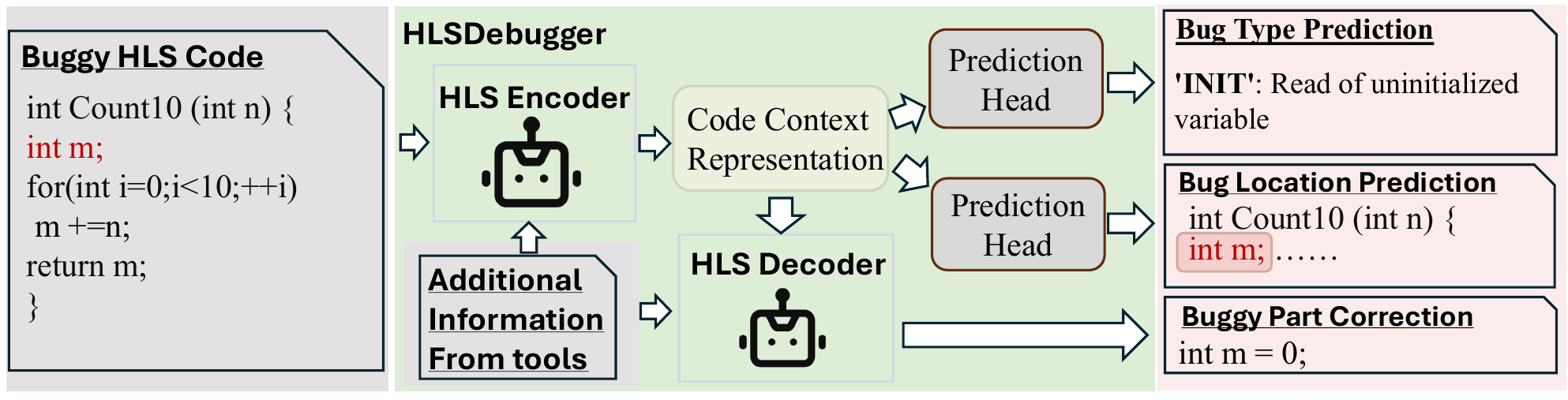}
    \label{fig:HLSDebugger-Solution}
    }
    \vspace{-.05in}
    \caption{Comparison between traditional LLM solutions on debugging tasks and our customized debugging structure on HLS code. Our solution in (b) is faster, more accurate, and offers improved performance robustness. 
    }
    \label{fig:comparision-framework}
    \vspace{-0.5cm}
\end{figure}

In recent years, the emergence of Large Language Models (LLMs) has provided unprecedented opportunities for automating code debugging tasks. Despite extensive debugging explorations in the software domain~\cite{yang2024cref,qin2024agentfl, lee2024unified, xu2024aligning, levin2024chatdbg, zhong2024ldb}, LLM solutions are almost unexplored in HLS, one of the most widely adopted agile design methodologies in the hardware domain. 
We believe LLM-assisted HLS debugging will help further boost the hardware design efficiency and narrow the productivity gap between software and hardware domains. 
Targeting LLM-assisted HLS debugging, particularly for \emph{logic bug} identification and correction, we have identified three key challenges that need to be addressed.  \looseness=-1

\textbf{Challenge 1. Circuit Data Scarcity.} The proprietary nature of circuit data and its intrinsic value in the semiconductor industry have caused a long-lasting barrier to circuit data availability. As a result, collecting sufficient designs in HLS code format is exceptionally difficult, thus preventing the training of LLM solutions for HLS debugging. The work of~\cite{10473893} introduces a benchmark for LLM-based identification of logic bugs in HLS. Targeting the data scarcity challenge, in this work, we generate and release a $25 \times$ larger-scale dataset with around 300K samples to support LLM training. This is a labeled dataset that supports supervised fine-tuning of LLMs, with each data sample including buggy code, corrected code, and detailed bug information.

\textbf{Challenge 2. Debugging Logic Bugs.} Many prior works on hardware and software debugging~\cite{tsai2023rtlfixer, yao2024hdldebugger,xu2024aligning,levin2024chatdbg, yang2024cref} focused on syntax bugs, which are easier to detect and correct based on corresponding compilers. However, logic bugs, where the code functionality diverges from the desired behavior, are much more challenging to handle. Compilers are not guaranteed to detect most logic bugs, necessitating more sophisticated functionality-aware methods to capture logic bug information. In this work, we target 8 types of known logic bugs that commonly exist in problematic HLS code, as suggested in the work of~\cite{10473893}. Their detailed information will be listed in Section~\ref{sec:setup}.

In the HLS domain, crafting reliable test cases for every HLS code sample has been labor-intensive, requiring significant domain expertise. The growing complexity of modern hardware makes hand-crafting enough test cases increasingly challenging, especially at such an early design stage. In this work, we focus on logic bugs of HLS code, without assuming any additional input such as test cases. As a result, both the identification and correction of bugs by LLMs are required, as we will introduce in the next subsection.

\begin{table*}[!t]
\centering
\vspace{-.2in}
\resizebox{0.94\textwidth}{!}{
\begin{tabular}{|c|c|c|c|c|c|c|c|c|}
\cline{1-9}
\multirow{2}{*}{Works} & \multirow{2}{*}{Domain} & Logic & New Training & New & Not Require & 
\multicolumn{3}{c|}{\text{Multi-task}} \\
\cline{7-9}
 & & Error & Dataset & Model & Test Cases & Bug Location & Bug Correction & Bug Info \\
\hline
RTLFixer \cite{tsai2023rtlfixer} & \multirow{3}{*}{RTL} &   &   &  &  &   & \textcolor{black}{\Checkmark} &  \\ 
HDLdebugger \cite{yao2024hdldebugger} &    &   &  Closed-source & Closed-source &  &   &  \textcolor{black}{\Checkmark} & \textcolor{black}{\Checkmark}\\ 
LLM4SecHW \cite{fu2023llm4sechw} &    & \textcolor{black}{\Checkmark} & Closed-source & Closed-source & \textcolor{black}{\Checkmark} & \textcolor{black}{\Checkmark}  & \textcolor{black}{\Checkmark}  & \textcolor{black}{\Checkmark}\\
\cline{1-9}
FixAgent \cite{lee2024unified} & \multirow{6}{*}{Software} & \textcolor{black}{\Checkmark} &  &  & \textcolor{black}{\Checkmark} & \textcolor{black}{\Checkmark} & \textcolor{black}{\Checkmark} & \textcolor{black}{\Checkmark}\\
D4C \cite{xu2024aligning} &   &  
\textcolor{black}{\Checkmark} 
&   &  
&   &  & \textcolor{black}{\Checkmark}  &  \\
AgentFL  \cite{qin2024agentfl} &   & \textcolor{black}{\Checkmark}  &   &  &
 & \textcolor{black}{\Checkmark} &   &  
\\
ChatDBG \cite{levin2024chatdbg} &   &  \textcolor{black}{\Checkmark} &   & &
&  & \textcolor{black}{\Checkmark} & \textcolor{black}{\Checkmark} \\
CREF \cite{yang2024cref} &   &  \textcolor{black}{\Checkmark} &   & 
&   &   & \textcolor{black}{\Checkmark}   & \textcolor{black}{\Checkmark} \\
LDB \cite{zhong2024ldb} &   &  \textcolor{black}{\Checkmark}  &   &
 &  &   & \textcolor{black}{\Checkmark}  & \textcolor{black}{\Checkmark}\\
\hline
\hline
\multirow{2}{*}{\textbf{HLSDebugger}} & \multirow{2}{*}{HLS} & \multirow{2}{*}{\textcolor{black}{\Checkmark}} & \multirow{2}{*}{Open-source} & \multirow{2}{*}{Open-source} &
\multirow{2}{*}{\textcolor{black}{\Checkmark}} & \multirow{2}{*}{\textcolor{black}{\Checkmark}} & \multirow{2}{*}{\textcolor{black}{\Checkmark}} & \multirow{2}{*}{\textcolor{black}{\Checkmark}}\\
      & & & & & & & & \\
\cline{1-9}
\end{tabular}
}
\caption{Comparison between HLSDebugger and related LLM-based \emph{debugging} works, all in other domains.} 
\label{tab:comparison-of-works}
\vspace{-.1in}
\end{table*}

\textbf{Challenge 3. Multi-Tasking in HLS Debugging.} Debugging HLS code without additional information (e.g., without test cases) typically requires a locate-then-correct process, making this process inherently multi-tasking. But this two-step process is challenging and can lead to error accumulation~\cite{xu2024aligning}. 
For example, when asked to identify the bug in the given code in the first step, traditional generative LLMs might wrongly produce an unrelated code snippet that is actually not even found in the original code. Such mistakes will make correcting bugs in the second step impossible.  
In addition, the traditional generative objective of LLMs (i.e., making generated output exactly match the label) is not sufficient for such multi-tasking debugging tasks.



In this paper, we propose a new solution named HLSDebugger, which tackles the aforementioned challenges. To the best of our knowledge, HLSDebugger is the first fine-tuned LLM solution performing both the identification and correction of logic bugs in HLS code. HLSDebugger first generates and releases a large labeled dataset targeting HLS logic bugs. The HLSDebugger model adopts an encoder-decoder structure, performing bug location identification, bug type prediction, and bug correction with the same model. 


\textbf{Dataset Generation.} We first address the circuit data gap by generating a large open-source dataset to train customized open-source LLMs on HLS debugging tasks. 
This data set contains around 300K data samples\footnote{Each data sample consists of a training instance containing a buggy code sample that begins with a module definition and concludes with the "end module" keyword, accompanied by corresponding bug indication and correction labels.} (each buggy sample will be used as a training sample), which is approximately 25 times the size in \cite{10473893}. 
It is also more than ten times larger than representative open-source hardware code datasets~\cite{liu2023rtlcoder, zhang2024mg} for other LLM-assisted hardware design tasks (e.g., RTL code generation). 
We generate the dataset by first collecting unlabeled data with mostly correct code, then subsequently inserting common bugs to generate corresponding buggy code.



\textbf{Encoder-Decoder Debugging Structure.}  To better adapt the LLM structure for this multi-tasking debugging application, we propose to utilize the encoder-decoder LLM structure with a customized loss function. As Figure \ref{fig:HLSDebugger-Solution} summarizes, the LLM encoder performs bug identification, both pinpointing bug location within the original code context and predicting the bug type. 
%
The decoder then incorporates the intermediate representation from the encoder to generate the bug correction. 
Experiments demonstrate that integrating bug location identification and correction in a unified encoder-decoder structure boosts the model performance in both tasks.

\textbf{Debugger Training Scheme.} Furthermore, we propose a new explicit training scheme for the customized encoder-decoder structure. This training scheme combines both \emph{bug identification loss} and \emph{bug correction loss} to train both the encoder and decoder parts of HLSDebugger simultaneously. 



Our contributions in HLSDebugger can be summarized as follows.

\begin{itemize}
    \item To the best of our knowledge, HLSDebugger provides the first fine-tuned LLM solution for both identification and correction of logic bugs in HLS code. 
    \item We generated and released an open-source, high-quality dataset for HLS debugging. Our dataset demonstrates effectiveness in improving existing open-sourced LLMs to outperform GPT-4 by more than $2\times$ in bug correction. 

    \item We proposed a customized encoder-decoder structure for HLS logic bug identification and correction. After fine-tuned with our dataset, it significantly outperforms the commercial GPT-4 by more than $3\times$ in bug correction. 
    
    \item We meticulously designed a customized training scheme for the encoder-decoder structure. This training scheme enables HLSDebugger's multi-tasking debugging process.

\end{itemize}

\begin{figure*}
    \centering
    \includegraphics[width=0.9\textwidth]{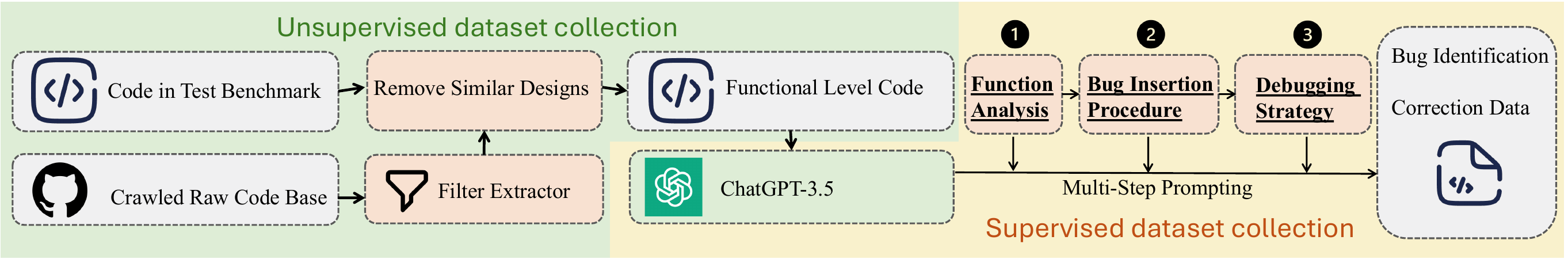}
    \vspace{-.05in}
    \caption{Our proposed dataset generation flow in HLSDebugger. It consists of two parts: 1) {unsupervised dataset collection} and 2) {supervised bug dataset generation}. The second supervised dataset generation process consists of three sub-steps:  \protect\circled{1} 
    \textbf{Function analysis}, \protect\circled{2} 
    \textbf{Bug insertion procedure}, \protect\circled{3} \textbf{Debugging strategy analysis}.} 
    \label{fig:Data-Flow}
    \vspace{-.1in}
\end{figure*}

\section{Related Work}\label{sec:pre}


We introduce related LLM for debugging solutions, covering software, RTL, and HLS in Section~\ref{subsec:llm-software}, \ref{subsec:llm-rtl}, and~\ref{subsec:llm-hls}, respectively. Table~\ref{tab:comparison-of-works} summarizes these related debugging works.

\subsection{LLMs for Software Debugging}
\label{subsec:llm-software}


In the software domain, benchmarks such as QuixBugs \cite{quixbugs} and CodeFlows \cite{CodeFlow} have been proposed for assessing model debugging capabilities. Most software debugging works \cite{xu2024aligning, levin2024chatdbg, yang2024cref,zhong2024ldb} do not undertake explicit bug location identification tasks. Instead, they assume test cases (e.g., unit tests) are available for each potentially buggy software code. As a result, they can directly utilize the information from test cases as known bug indicators to guide bug correction. For similar reasons, most software debugging works do not strictly distinguish logic and syntax bugs.

In this work, we focus on a more challenging scenario where the logic bugs can not be identified directly by the compiler, and no test cases are provided~\cite{10473893}. LLMs are tasked with identifying and correcting bugs at the same time.

\subsection{LLMs for RTL Debugging}
\label{subsec:llm-rtl}


In the hardware domain, most works focus on RTL instead of HLS code. The works of \cite{yao2024hdldebugger, tsai2023rtlfixer, fu2023llm4sechw, ma2024verilogreader, qiu2024explaining} target debugging of RTL. Works~\cite{yao2024hdldebugger, tsai2023rtlfixer} build robust databases to support LLM debugging using RAG, which needs error information from test cases.
LM4SecHW~\cite{fu2023llm4sechw} created a dataset from GitHub for fine-tuning LLMs by gathering commit information tagged with `bug' and `enhancement' from open-source hardware repositories. This dataset enhances LLM's capabilities in identifying and correcting function-related bugs, flaws, or possible improvements. However, such ``version-control" data from GitHub does not ensure that a subsequent commit actually corrects bugs in prior code.

In our work, we are facing a more challenging scenario where the location of the bug is not available. We compared different LLMs in HLS debugging without providing location information and observed a significant drop in debugging performance. To address this challenge, we propose the HLSDebugger that simultaneously tackles both the location problem and the bug-fixing problem.

\subsection{LLM for HLS Debugging \& Generation}
\label{subsec:llm-hls}

Compared to RTL and software development, there is a lack of mature debugging tools and methodologies tailored for HLS. HLS operates at a higher level of abstraction, which can hide details that are crucial for hardware design but are not immediately visible in the source code, making it harder to pinpoint issues related to hardware implementation. Even though there are works like C2HLSC \cite{collini2024c2hlsc} to explore HLS tasks related to code generation, LLMs' solution on HLS debugging tasks remains untouched.

Wan et al. \cite{10473893} introduced an HLS dataset with around 12K data samples. This dataset is more appropriate for benchmarking various LLM solutions in HLS code debugging. This work does not contribute new models for HLS debugging. 
HLSPilot \cite{Xiong2024HLSPilotLH} targets HLS generation tasks, emphasizing performance bottlenecks and co-design for CPU and FPGA. It prompts GPT to create a framework but does not offer open-source models.

\section{Problem Formulation}
\label{sec:problem-formulation}

We first formulate the HLS debugging problem. Given a potentially buggy code sample $C_\text{buggy}$ as input, we denote the actual bug (i.e., buggy snippet) in the input code sample as $s_\text{buggy}$, with $s_\text{buggy} \subseteq C_\text{buggy}$. This debugging process includes two tasks: 1) identify the bug snippet $s_\text{buggy}$ in the code sample $C_\text{buggy}$; 2) generate a corrected code sample $C_\text{correct}$. The expected outputs for each task during debugging are described below.    
\begin{itemize}
    \item \textbf{Bug Identification}. 
    Assume there are altogether $N$ tokens in the given code sample $C_\text{buggy}$. The label of bug identification can be represented as a binary list $\{I_1, I_2, ..., I_N\}$, where each $I_i$ is a binary indicator denoting whether the $i^{th}$ token of $C_\text{buggy}$ is part of the buggy snippet $s_\text{buggy}$. This can also be expressed as $s_\text{buggy} = \{C_\text{buggy}[I_i] \,|\, I_i == 1\}$.    
    \item \textbf{Bug Correction.} After identifying the buggy tokens, the LLM needs to generate the corrected code sample by replacing the buggy snippet $s_\text{buggy}$ in input code sample $C_\text{buggy}$. The correct code sample is denoted as $C_\text{correct}$. We denote the corresponding corrected bug snippet as $s_\text{correct}$, such that $s_\text{correct} \subseteq C_\text{correct}$.  
\end{itemize}
Our goal is to develop an LLM solution, denoted as $LLM$, to maximize the accuracy of the process below: 
\begin{align*}
    LLM( C_\text{buggy}) \rightarrow  \{I_1, I_2, ..., I_N\}, C_\text{correct}
\end{align*}

\section{Methodology}\label{sec:method1}


This Section consists of three sub-sections corresponding to our major contributions: 1) Dataset generation; 2) Encoder-decoder structure; 3) Encoder-decoder training scheme. 

\subsection{Dataset Generation}
\label{sec:data-gen}
To address the data scarcity problem for fine-tuning LLMs on HLS debugging, we proposed a dataset generation procedure and generated a large labeled dataset with approximately 300K data samples. 
This supervised dataset includes buggy HLS code sample $C_\text{buggy}$ and detailed bug labels. The bug label provides information for both bug identification and correction: buggy snippet $s_\text{buggy}$, its bug type $T$, and corresponding correct code snippet $s_\text{correct}$.  

As Figure~\ref{fig:Data-Flow} shows, the dataset generation process consists of two main parts: 1) Unsupervised dataset collection; and  2) Supervised bug dataset generation. We first collected an unsupervised dataset with correct HLS code from GitHub and GPT-3.5 \cite{achiam2023gpt} generations. Based on this initial unlabeled dataset, we generate the supervised bug dataset by inserting bugs to the original code through the CoT process by GPT-3.5 \cite{achiam2023gpt}. As introduced, the final supervised dataset provides buggy code $C_{buggy}$, correct code $C_{correct}$, and bug information for each data sample.

\textbf{Unsupervised Dataset Collection.}
First, we crawled a substantial number of HLS codes from GitHub\footnote{The GitHub projects are collected under the HLS Topic (\url{https://github.com/topics/hls} with around 100 projects.)}, and then we applied a filtering mechanism to remove noisy notation and markdown files.
These crawled data constitute the initial unsupervised dataset with around 30K code samples\footnote{We split the crawled context into individual code samples, each starting with an ``module" and ending with an ``end module" keyword.}. To further augment the dataset, we converted many RTL codes from another RTL dataset proposed by \cite{liu2023rtlcoder} into HLS codes using GPT-3.5 \cite{achiam2023gpt}\footnote{This augment process is mainly intended to provide a larger code context for training open-source LLMs, thus functional matching between the original RTL code and converted HLS codes is not strictly enforced.}. In this augmentation step, we further generated around 10K additional code samples. We subsequently prompted GPT-3.5 \cite{achiam2023gpt} to refine the potential flaws or errors in both crawled and augmented code samples (Each code sample is denoted as $C_\text{correct}$).

\textbf{Supervised Bug Dataset Generation}. Based on the collected unsupervised dataset from the first step, we perform bug insertion to build our desired supervised dataset for fine-tuning LLMs. This bug insertion process begins by pinpointing potential bugs that can be inserted into each original correct code sample $C_\text{correct}$. After the analysis, we create various buggy code samples $C_\text{buggy}$ by substituting originally correct snippet $s_\text{correct}$ in $C_\text{correct}$ with buggy snippet $s_\text{buggy}$.  


As introduced, this supervised dataset will include buggy HLS code and detailed bug labels.  We prompt the GPT-3.5 \cite{achiam2023gpt} model with CoT to generate the supervised bug dataset through three sub-steps: 1) function analysis; 2) bug insertion procedure; and 3) debugging strategy analysis. Each generation step takes the output of its previous step as its input.


\begin{figure}[!t]
\centering
\includegraphics[width=0.5\textwidth]{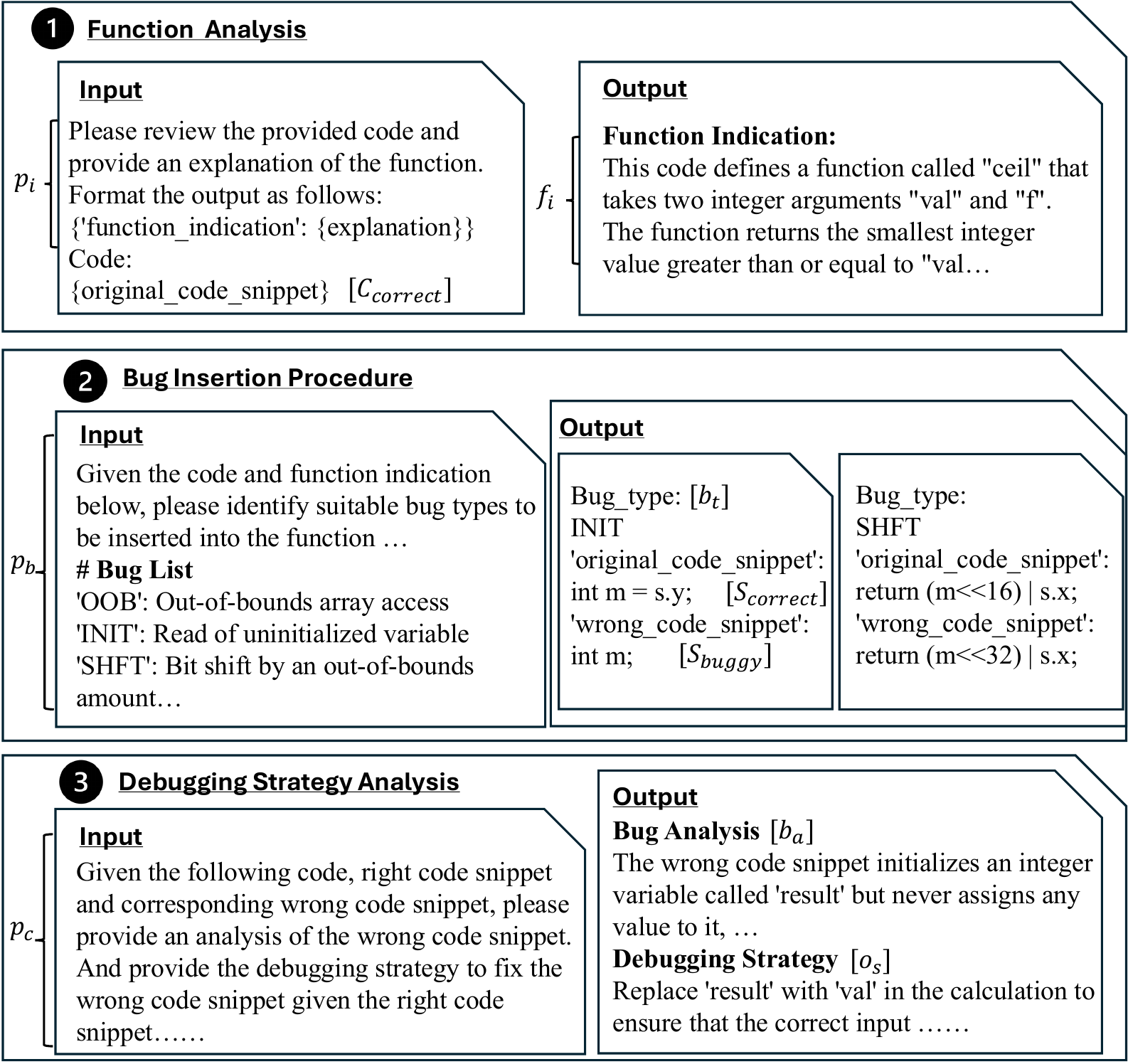}
\caption{Our supervised dataset creation process involves three steps: (a) \textbf{Function analysis:} LLM reviews and refines code to generate functionality indication $f_i$. (b) \textbf{Bug insertion procedure:} LLM selects and inserts bugs based on a provided list and function indication, generating a dataset for bug identification and correction. (c) \textbf{Debugging strategy analysis:} A subset of data is used to generate bug analysis and debugging strategies, enhancing LLM's bug correction capabilities. }
\label{fig:multi-step-prompt}
\end{figure}

Figure~\ref{fig:multi-step-prompt} (a) shows the step 1 named function analysis. We first design an instruction prompt $p_i$ and concatenate the correct code sample $C_\text{correct}$ after it. 
We then feed the content $(p_i || C_\text{correct})$ into the GPT-3.5 \cite{achiam2023gpt}, which we denote as $\mathbb{G}_1$. $\mathbb{G}_1$ is instructed to generate the function indication $f_i$: 
\begin{align*}
    f_i = \mathbb{G}_{1}(p_i || C_\text{correct})
\end{align*}

The output $f_i$ will be the input of the next step. 
As Figure~\ref{fig:multi-step-prompt} shows, the functionality indication $f_i$ involves the variable definitions, inputs, outputs, and operations.




Figure~\ref{fig:multi-step-prompt} (b) shows the step 2 named bug insertion procedure. 
GPT-3.5 ($\mathbb{G}_2$) \cite{achiam2023gpt} is prompted by $p_b$ to select relevant bug types suitable for the given refined code $C_\text{correct}$ from 18 types of logic bugs, with details given in Section~\ref{sec:setup}. The bug list will be directly concatenated into the prompt $p_b$. Each resulting bug sample contains the bug type $b_t$, original code snippet (correct code snippet) $s_\text{correct}$, and buggy code snippet $s_\text{buggy}$.  
 This process can be formulated as follows:
\begin{align*}
    \{s_\text{correct}, s_\text{buggy}, b_t\} = \mathbb{G}_{2}(p_b || C_\text{correct} || f_i )
\end{align*}
We create the buggy code sample $C_\text{buggy}$ simply by replacing the original correct snippet $s_\text{correct}$ with the buggy snippet $s_\text{buggy}$. The main supervised dataset is generated through these first two steps, including 300K data samples.



Figure~\ref{fig:multi-step-prompt} (c) shows the step 3 named the debugging strategy analysis.  
We further create a small dataset, including bug analysis $b_a$ and debugging strategy $o_s$. 
For the generation, we combine the specific prompt $p_c$,  buggy code snippet $s_\text{buggy}$, which pinpoints the exact bug, and the corresponding correct code snippet $s_\text{correct}$.  
\begin{align*}
    b_a, o_s = \mathbb{G}_{3}(p_c || C_\text{buggy} || s_\text{correct} || s_\text{buggy})
\end{align*}


The bug analysis $o_s$ explains the cause of the bug in the buggy snippet $s_\text{buggy}$, indicating its difference from the correct functionality. The debugging strategy $o_s$ provides modification suggestions based on the bug analysis $b_a$ and the buggy sample $C_\text{buggy}$ and correct sample $C_\text{correct}$.



The final supervised dataset includes both the main dataset $\{s_\text{correct}, s_\text{buggy}, b_t\}$ from step 2 and additional debugging strategy information from $\{b_a, o_s\}$ step 3, as formulated below.
\begin{equation}
       \{s_\text{correct}, s_\text{buggy}, b_t\}, \{b_a, o_s\} = \mathbb{G}_{3}\,|\,\mathbb{G}_{2}\,|\,\mathbb{G}_{3} (p_i || C_\text{correct})
\end{equation}

%

\begin{figure}[!t]
\centering
\includegraphics[width=0.5\textwidth]{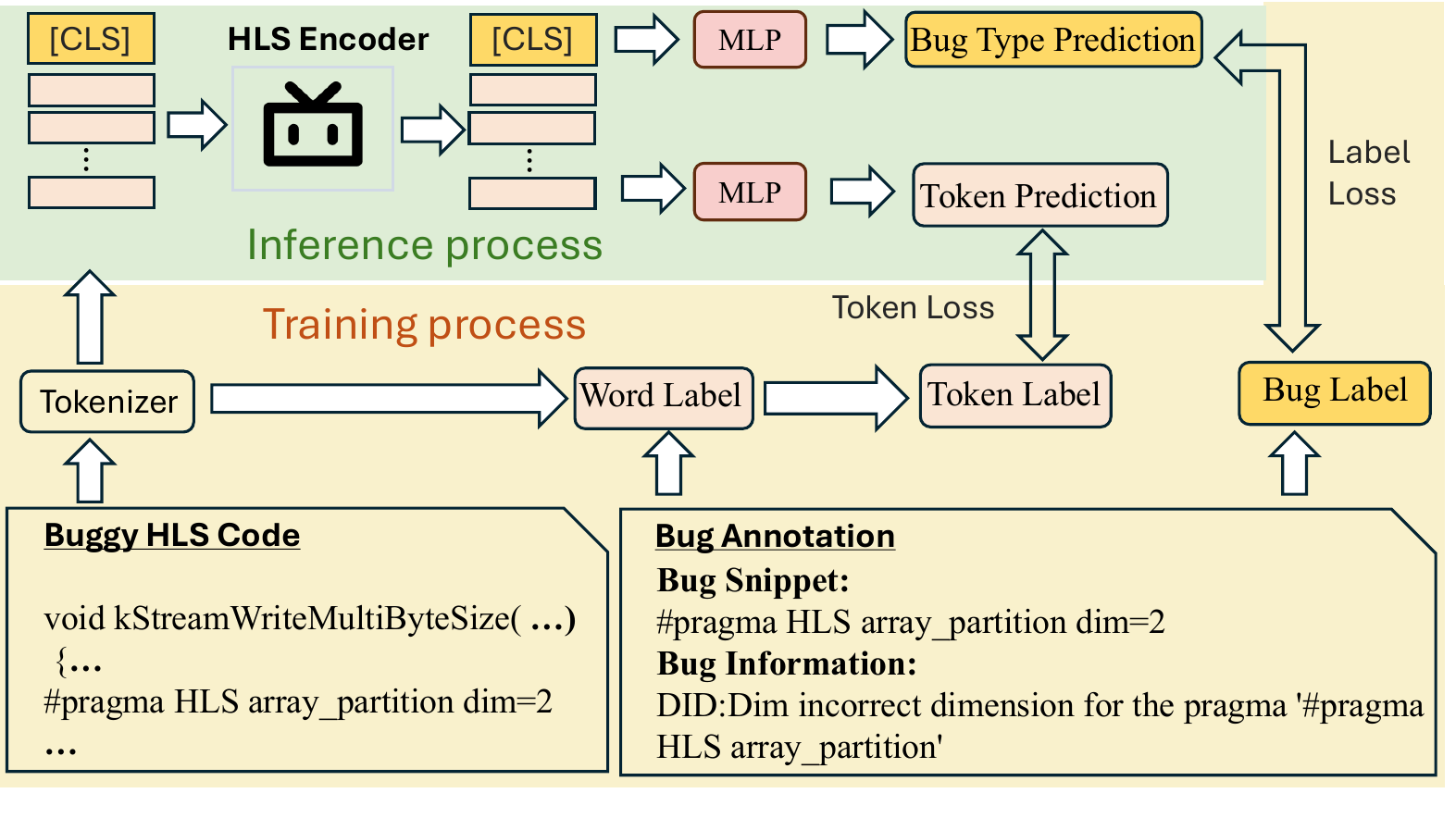}
\vspace{-.1in}
\caption{The {inference} and {training} schematic for HLS Encoder. Given the tokenized HLS buggy code, we begin by concatenating the [CLS] token at the beginning. The [CLS] token represents the code context, which will later be used by the MLP head to predict the bug type. }
\label{fig:encoder-training}
\end{figure}

\subsection{Encoder-Decoder Structure of HLSDebugger}
\label{sec:encoder-decoder}


In this section, we introduce the encoder-decoder structure of HLSDebugger, illustrated in Figure~\ref{fig:HLSDebugger-Solution}. Our preliminary studies indicate that large language models (LLMs) frequently misidentify buggy code locations, leading to incorrect corrections. We define this issue as error accumulation, which presents a significant bottleneck in the traditional single decoder-only LLM approach during the localization-then-correction process.
Additionally, even when they identify the correct snippets, they frequently fail to produce appropriate corrections. This might be because the LLMs still fail to fully understand the context and bug reason.
To address this, we propose an encoder-decoder architecture. This approach allows us to train the encoder explicitly to identify bug locations more accurately. Additionally, the encoder provides a soft representation of the bug location, rather than a potentially misleading explicit string. This soft representation, integrated with the decoder through cross-attention, includes more contextual information beneficial for accurate bug correction, thereby enhancing the overall performance of the system.

\begin{figure}[!t]
\centering
\includegraphics[width=0.5\textwidth]{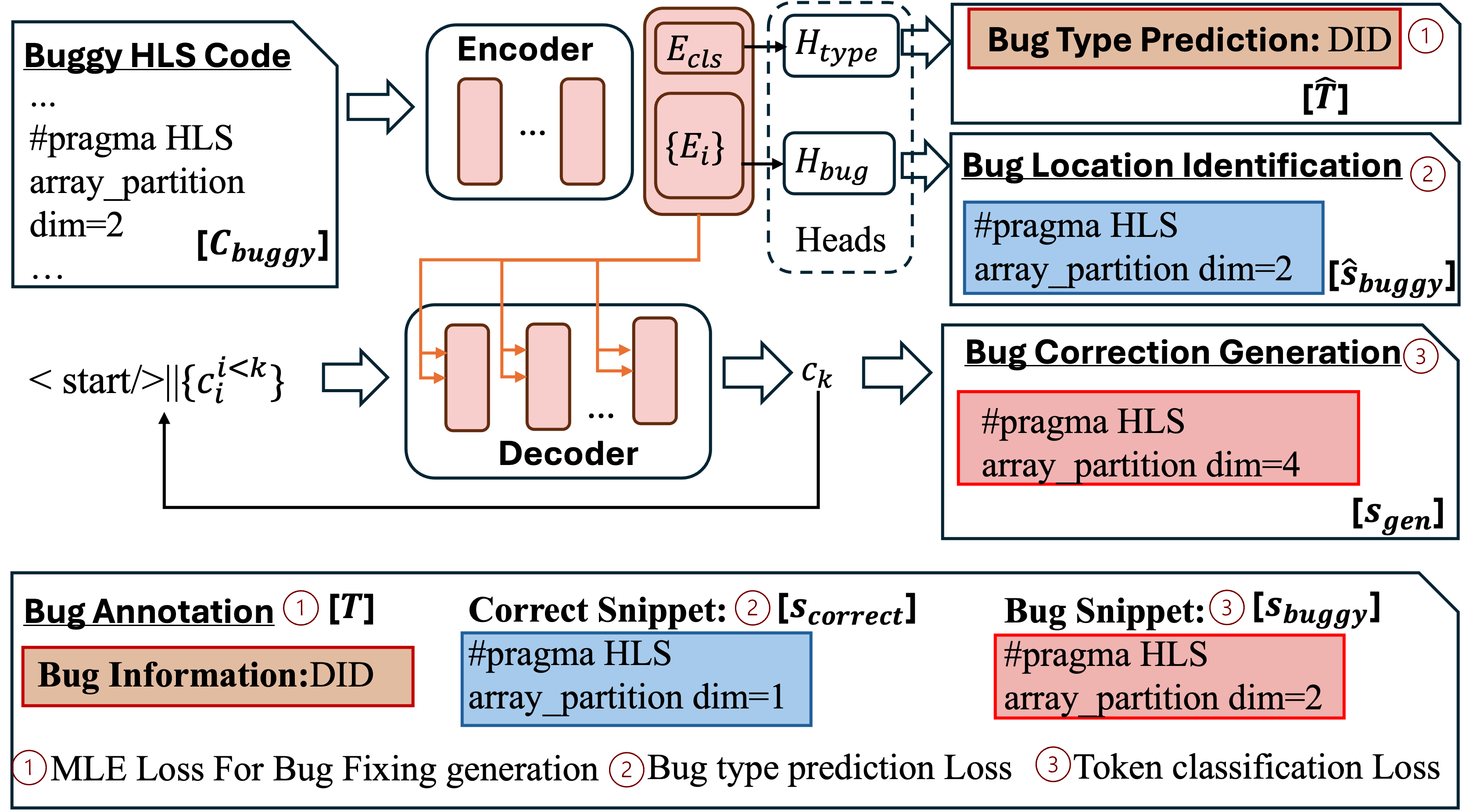}
\caption{The interaction between the HLS encoder and HLS decoder within HLSDebugger, along with the ensemble combined loss. The total loss function is the sum of the 
bug type prediction 
loss, bug location loss, and bug correction loss.} 
\label{fig:ensemble-training}
\vspace{-.1in}
\end{figure}

Based on this observation, we propose to adopt an encoder-decoder structure for HLSDebugger, which consists of an HLS Encoder and an HLS Decoder, as shown in Figure~\ref{fig:ensemble-training}. The HLS Encoder is tasked to identify the location and type of bugs. It also provides contextual information to support the HLS Decoder by directing their attention to the critical buggy snippet within the context.

\textbf{HLS Encoder.} 
Figure \ref{fig:encoder-training} showcases the process of the HLSEncoder generating the context representation and using the prediction head for bug type prediction and bug location identification. 
The HLSEncoder uses a stack of transformer blocks, each consisting of self-attention layers and feed-forward networks. The self-attention mechanism allows the encoder to capture the context and relationships between different parts of the input sequence. The representation generated by the HLSEncoder is a high-dimensional vector, which encapsulates its semantic meaning. 
This representation is then used as input to two distinct heads: one for predicting the type of bug and another for pinpointing its location.   
Assume there are $N$ tokens in the buggy code $C_\text{buggy}$. The input into the encoder is the concatenation of the $N$-token buggy code $C_\text{buggy}$ after a special token named $[CLS]$, denoted as $ ([CLS] || C_\text{buggy})$. This special leading token $[CLS]$ represents the whole code functionality as well as the bug information. The encoder will generate an embedding vector for each of the $N$ tokens, denoted as $\{E_1, E_2, ..., E_N \}$, as well as a leading embedding $E_\text{cls}$ for the special token $[CLS]$. This encoding process is formulated below.    
\begin{align}\label{formu:HLS-Encoder}
    E_\text{cls},~ E_1, ~..., ~E_N = \operatorname{Encoder}(\text{[CLS]} || C_\text{buggy})
\end{align}  

After encoding the potentially buggy code $C_\text{buggy}$, these context embeddings $E_\text{cls}$, $\{E_1, E_2, ..., E_N \}$ will be fed into the \emph{prediction heads} ($\mathbb{H}_\text{type}, \mathbb{H}_\text{bug}$). Each prediction head is based on a basic multilayer perceptron (MLP). Prediction heads $\mathbb{H}_\text{type}$ and $\mathbb{H}_\text{bug}$ will predict the bug type (prediction denoted as $\hat{T}$) of the overall code and identifying whether each of the $N$ tokens is buggy (prediction denoted as $\hat{I}_1, ..., \hat{I}_N$), respectively:
\begin{align*}
    \hat{I}_1, ..., \hat{I}_N &=  \mathbb{H}_\text{bug}(E_1, ..., E_N) \\
    \hat{T} &= \mathbb{H}_\text{type}(E_\text{cls})
\end{align*}


\textbf{HLS Decoder.} The HLS Decoder is tasked to generate the bug correction $C_\text{gen}$, which is similar to the HLS Encoder as a set of transformer blocks, but it also includes cross-attention layers that allow it to `attend' to the encoded input representation. The final layer of the decoder generates sequences of corrected buggy snippets.
Figure~\ref{fig:ensemble-training} shows the dynamics between the encoder and decoder.
At each time, $\operatorname{Decoder}$ will generate the probability prediction for the next token $c_k$ in all vocabularies, based on all previous tokens before $c_k$, which can be denoted as $\{c_i^{\{i<k\}}\}$. 
\begin{equation}
    c_k = \operatorname{Decoder}(\text{$<$start/$>$}||\{c_{i}^{i<k}\}, \{E_\text{cls}, E_1, ..., E_N\})
\end{equation}
$c_k$ represents the $k_{th}$ token of the correct code snippet sequence $s_\text{gen}$. 
The input into Decoder $<$start/$>||\{c_i^{i<k}\}$ is the concatenation of the start token $<$start/$>$ and previous generated sequence. The other input $\{E_\text{cls}, E_{1}, ..., E_N\}$ is the output embeddings from the encoder. It is provided through the decoder's cross-attention layers \cite{vaswani2017attention}. In the cross-attention layers, the encoder outputs $\{E_\text{cls}, E_1, E_2, ..., E_N\}$ will be taken as key (K) and value (V) embeddings, while the intermediate result from each previous decoder layer will be treated as a query (Q). 
The interaction between decoder intermediate results and encoder output embeddings in cross-attention layers enables the decoder to selectively attend to relevant tokens and improve the quality of bug correction. 

The decoder and encoder are trained together to learn the interaction with each other with a customized loss, as we will introduce in the following sub-section.

\begin{figure}
    \centering
    \label{fig:2-phase-train}
    \includegraphics[width=0.4\textwidth]{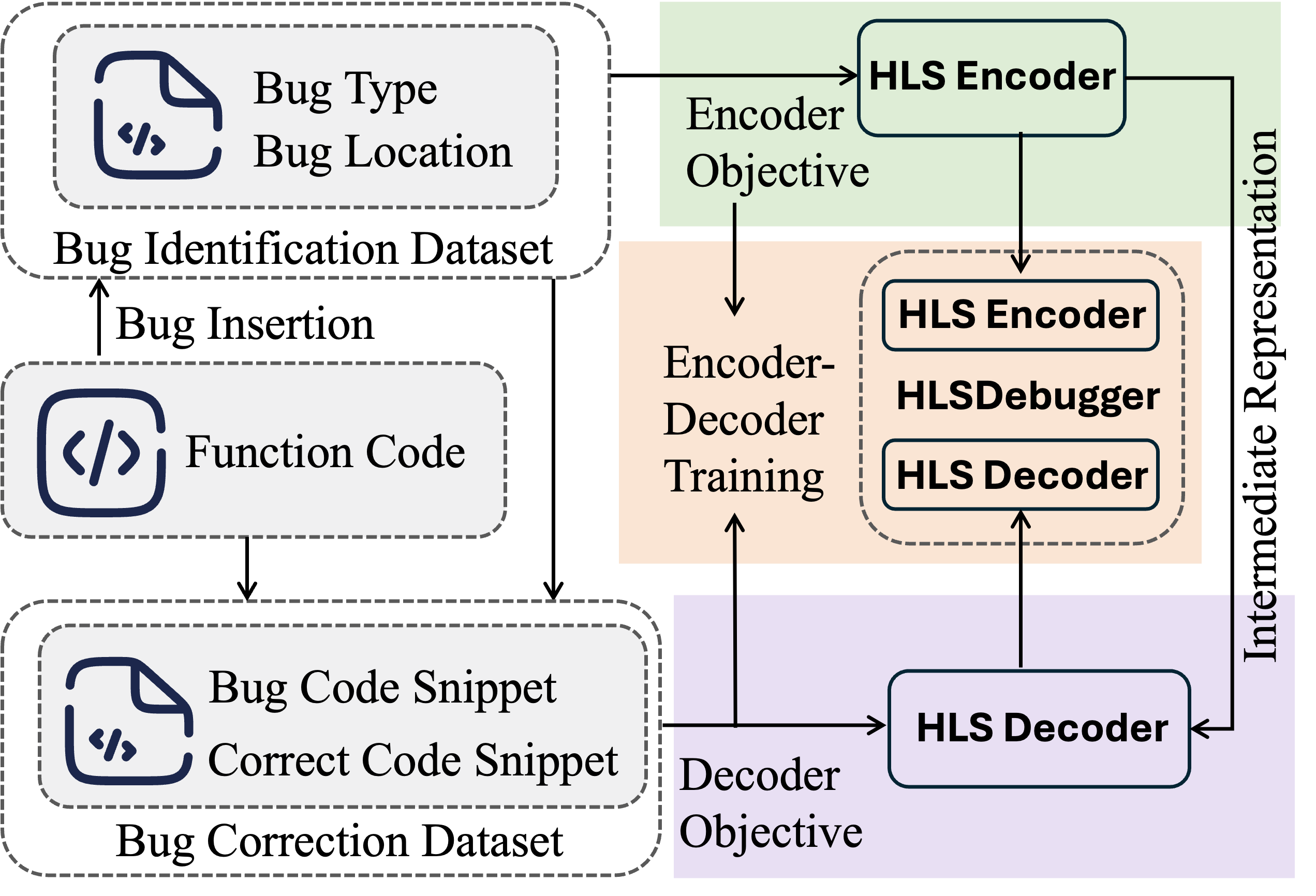}
    \caption{
    Combination of \emph{encoder objective} and \emph{decoder objective} for the encoder-decoder training scheme.
    }
    \label{fig:Two-phase-train}
   \vspace{-.2in}
\end{figure}

\subsection{Training Scheme for Encoder-Decoder Structure}
\label{sec:two-phase}

Most decoder-only LLMs for generative tasks adopt the basic maximum likelihood estimation (MLE) loss for their fine-tuning process. This cannot fulfill our requirements for fine-tuning encoder-decoder structured HLSDebugger on multiple tasks. We fill the gap by proposing a customized training scheme that simultaneously trains both the encoder and decoder with distinct loss functions, as illustrated in Algorithm~\ref{alg:phase-2-train}.

\RestyleAlgo{ruled}

\SetKwComment{Comment}{/* }{ */}

\textbf{Encoder Training Loss.} As introduced in Section \ref{sec:encoder-decoder}, the input to the encoder is the buggy code sample $C_\text{buggy}$, and the prediction based on encoder outputs is the bug type prediction $\hat{T}$ and the binary bug classification for each token $\hat{I}_1, ..., \hat{I}_N$. The encoder's objective is to minimize the combined loss of bug type prediction and bug location identification.

The bug type prediction loss is the cross-entropy loss (TypeLoss in Algorithm~\ref{alg:phase-2-train}) for multi-label classification on $\hat{T}$, denoted as $L_\text{type}$. The token classification loss for bug location identification is the simple binary cross-entropy loss (BugLoss in Algorithm~\ref{alg:phase-2-train}) on $\{\hat{I}_1, ..., \hat{I}_N\}$, denoted as $L_\text{{bug}}$. 
Additionally, the token distribution is unbalanced since most tokens are labeled with \emph{no bug} (i.e., false in token classification). To account for the imbalanced distribution, we assign a scaling factor of $\alpha_t$ to true labels and $\alpha_f$ to false labels when calculating $L_\text{{bug}}$. For the combinational loss of bug type and token classification, we further assign a factor $\alpha_\text{type}$ to bug type loss $L_\text{type}$ and a factor $\alpha_\text{bug}$ to token classification loss $L_\text{bug}$. The loss function of the encoder is defined below:
\begin{align}\label{formu:loss-phase-1}
    L_\text{encoder} = \alpha_\text{type} \cdot L_\text{type} + \alpha_\text{bug} \cdot L_\text{bug}
\end{align}

\textbf{Decoder Training Loss.} During training, the correct code snippet $s_\text{correct}$ in the desired code sample $C_\text{correct}$ serves as the ground truth for the decoder. $s_\text{correct}$ is a sequence of tokens with a total length of $K$. The decoder's output, $s_\text{gen}$, is compared against $s_\text{correct}$ to calculate the loss.
Denote the sequence of tokens before $k_{th}$ in $s_\text{correct}$ as $\{c_{i}^{i<k}\}$, the overall MLE loss is:
\begin{equation}
    L_\text{decoder} =  - \sum_{k=1}^{K} \text{log} \operatorname{Decoder}(\text{$<$start/$>$}||\{c_{i}^{<k}\})
\end{equation}
In this equation, the output from the HLS Decoder is the probability of the $k_{th}$ token across all vocabularies.
The loss function aims to maximize the log-likelihood of the correct snippet sequence $s_\text{correct}$, encouraging the decoder to predict the correct token at each step in sequence generation.

\textbf{Combined Loss for Ecoder-Decoder.} Finally, we combine the loss terms of both encoder and decoder, scaling them with hyper-parameters $\alpha_\text{encoder}$ and $\alpha_\text{decoder}$. The final loss is shown below.    
\begin{align}
L_\text{all} = \alpha_\text{encoder} \cdot L_\text{encoder} + \alpha_\text{decoder} \cdot L_\text{decoder}  \label{formu:loss-phase-2}  
\end{align}


\begin{algorithm}[!t]\small
\caption{Training for Encoder-Decoder Structure } \label{alg:phase-2-train}
\textbf{Intialize Training}:
    training epochs: $N$,
    training dataset: $D$,
    set iteration index $iter = 0$,
    optimizer: $\textbf{Optm}$\;
    
    \textbf{Set Loss Factor}:
    Bug type: $\alpha_\text{type}$, bug location: $\alpha_\text{bug}$, buggy token factor: $\alpha_t$, correct token factor: $\alpha_f$,
    bug correction generation $\alpha_\text{decoder}$
    \;
    
    $\textbf{TypeLoss} \gets \text{LogitCrossEntropyLoss}$\;
    
    $\textbf{BugLoss} \gets \text{LogitCrossEntropyLoss}(\alpha_t, \alpha_f)$\;
    
    $\textbf{GenLoss} \gets \text{LogitMaximumLikelihoodEstimation}$\;
    \While {iter $< N$}{
        \For {batch $B$ in $D$}{
        $C_\text{buggy}, \{I_1, I_2, ..., I_N\}, T, s_\text{correct} \gets B$\;
        
        $E_\text{cls}, \{E_1, E_2, ..., E_N\} = \text{Encoder}(\text{[CLS]} || {C_\text{buggy}})$ \;
        \{$\hat{I}_1, ..., \hat{I}_N\} = \mathbb{H}_\text{bug}(\{E_{1}, E_2, ..., E_N\})$  \;
        $\hat{T} = \mathbb{H}_\text{type}(E_\text{cls})$ \;
        ${s}_\text{gen} \leftarrow \text{Decoder}(E_\text{$<$start/$>$}, \{E_\text{cls}, E_1, ..., E_N\})$\;
        $L_\text{type}  = \textbf{TypeLoss}(T, \hat{T})$ \;
        $L_\text{bug} = \textbf{BugLoss}(\{I_1, ..., I_N\},\{\hat{I}_1, ..., \hat{I}_N\})$ \;

        $L_\text{encoder} = \alpha_\text{type} \cdot L_\text{type} + \alpha_\text{bug} \cdot L_\text{bug}$
        
        $L_\text{decoder} = \textbf{GenLoss}(
        s_\text{correct}, {s}_\text{gen}
        )$\;

        $L_\text{all} = \alpha_\text{encoder} \cdot L_\text{encoder} + \alpha_\text{decoder} \cdot L_\text{decoder}$ \;
        
        $L_\text{all}.\operatorname{backward}()$ \;
        $\textbf{Optm}.\operatorname{step}()$\;
        
        }
    }
\end{algorithm}

\begin{table*}[]
    \vspace{-.2in}
    \centering
    \begin{tabular}{|c||c|c|c|c||c|c|c|c||c|c|}
    \hline
        Model & \multicolumn{4}{c||}{Token-wise accuracy} & \multicolumn{4}{c||}{Line-wise accuracy} & \multicolumn{2}{c|}{Code-wise accuracy}  \\
        \hline
        & Precision & Recall & f1 & RoC &  Precision & Recall & f1 & RoC &  Top-1 & Top-5 \\
        \hline
        \hline
         GPT-3.5 \cite{achiam2023gpt}& 10.1\% & 29.1\% & 0.136 & 0.513 & 10.6\% & 15.4\% & 0.121 & 0.554 & 16.3\% & 53.4\% \\
        GPT-4 \cite{achiam2023gpt} & 12.2\% & 37.1\% & 0.168 & 0.554 & 13.3\% & 17.5\% & 0.145 & 0.566 & 19.7\% & 54.7\% \\ 
        Gemma-7B \cite{team2024gemma} & 7.2\% & 21.0\% & 0.097 & 0.447 & 7.6\% & 22.3\% & 0.104 & 0.451 & 15.5\% & 60.7\% \\
        \hline
        Gemma-7B UFT & 7.3\% & 18.8\%  & 0.092 & 0.476 & 7.5\% & 19.6\% & 0.096 & 0.475 & 26.1\% & 67.3\% \\
        Gemma-7B SFT & 22.2\% & 47.7\% & 0.279 & 0.648 & 20.0\% & 46.4\% & 0.255 & 0.632 & 37.1\% & 69.9\% \\
        \hline
        \hline
        HLSEncoder & 46.1\% & 79.8\% & 0.539 & 0.913  & 50.5\% & 75.4\% & 0.564 & 0.899 & 76.7\% & 86.9\% \\
        \textbf{HLSDebugger} & 54.3\% & 83.5\% & 0.614 & 0.935 & 57.7\% & 78.4\% & 0.627 & 0.923 & 79.5\% & 87.6\% \\
        \hline
    \end{tabular}
    \caption{Evaluation of bug identification accuracies on different granularities.}
    \label{tab:bug-identification-overall}
    \vspace{-0.1in}
\end{table*}

During the training process, we noticed that bug correction by the decoder is more challenging than identifying bug types and locations by the encoder. Therefore, we attribute a higher loss scaling factor $\alpha_\text{decoder}$ to the decoder loss $L_\text{decoder}$. Based on the combined loss, the encoder and decoder in the HLSDebugger are trained simultaneously to learn a better interaction through cross-attention layers.

\section{Experimental Setup} \label{sec:expr-setup}

Before introducing results, in this section, we will introduce the details about the setup of our experiment, including the testing benchmark and tasks (Sec~\ref{sec:setup}), the implementation details of our HLSDebugger model (Sec~\ref{sec:implementation-detail}), the evaluation metrics (Sec~\ref{sec:sub-str-metric}), and the selection of baseline models (Sec~\ref{sec:baseline-select}).

\subsection{Experimental Benchmark and Task}\label{sec:setup}

\textbf{Evaluation Benchmark.} To evaluate HLSDebugger and other LLM solutions, we utilize the benchmark\footnote{To avoid a long overall inference time of LLMs during decoding, we randomly selected 2K samples from the benchmark in~\cite{10473893} for the evaluation of all models. Our experiments validate that all model testing accuracies converge very well after more than 500 samples are tested (i.e., change in average test accuracy is negligible when adding more test samples).} from \cite{10473893}, which comprises 8 different types of logic bugs.
All of the types of bugs and a broad spectrum of designs are covered in the selected test samples. The 8 types of bugs are listed as follows \cite{10473893}:
\begin{enumerate}
    \item \texttt{`OOB'} - Out-of-bounds array access
    \item \texttt{`INIT'} - Read of uninitialized variable
    \item \texttt{`SHFT'} - Bit shift by an out-of-bounds amount
    \item \texttt{`INF'} - An infinite loop arising from an incorrect loop termination
    \item \texttt{`USE'} - Unintended sign extension
    \item \texttt{`MLU'} - Errors in manual loop unrolling
    \item \texttt{`ZERO'} - Variable initialized to zero instead of nonzero initializer
    \item \texttt{`BUF'} - Copying from the wrong half of a split buffer
\end{enumerate}

\arrayrulecolor{black}

The bug location identification performance is evaluated in token-wise, line-wise, and code-wise granularities. The bug correction generation accuracy is evaluated by matching the ground truth correct code snippet with the generated outputs. Detailed metrics are introduced in the section (\ref{sec:sub-str-metric}).

\subsection{HLSDebugger Implementation in Experiment}
\label{sec:implementation-detail}

\textbf{HLSDebugger Structure Details.} We developed the HLSDebugger based on a pre-trained CodeT5-Large \cite{Wang2021CodeT5IU} model, with 24 transformer layers for both encoder and decoder (the dimensions of the embeddings are 1024, and the head dimension is 16). The number of layers for our MLP prediction heads ($\mathbb{H}_\text{bug}$, $\mathbb{H}_\text{type}$) is set to 3.

\textbf{HLSDebugger Training Details.} To avoid potential information leakage (i.e., overlap of data samples between training dataset and testing benchmark~\cite{10473893}), we utilize the Rouge-L \cite{lin-2004-rouge} metric to calculate the similarity between codes in our unsupervised dataset and that in the testing benchmark \cite{10473893}. We eliminate code samples that exhibit high similarity to the testing benchmark~\cite{10473893}. This operation removes approximately 2K code samples with a Rouge-L similarity greater than 0.5. The finalized number of training code samples in our dataset is around 300K. For training HLS Debugger, the factors $\alpha_\text{type}, \alpha_\text{bug}, \alpha_\text{gen}$ in the Algorithm~\ref{alg:phase-2-train} for are: 0.2, 2, 10.
The scaling factors $(\alpha_t, \alpha_f)$ for token identification loss (BugLoss in Algorithm~\ref{alg:phase-2-train}) are set as 0.05, 1.
We also fine-tuned an Encoder with only bug identification loss, denoted as HLSEncoder, for an ablation study. The factors are set with the same proportion ($\alpha_\text{type},~\alpha_\text{bug} = 0.2,~2$) but larger scale for better convergence.

\subsection{Evaluation Metrics} \label{sec:sub-str-metric}

In this subsection, we introduce the metrics used to evaluate the performance of LLMs for both bug identification and correction.

\textbf{Bug Identification Metrics.}\label{sec:identification-metric}
We evaluate the bug identification accuracy of all models at three different granularities: 1) Token-wise classification. Similar to the encoder training process, token-wise evaluation treats bug location as a token classification problem. 2) Line-wise string classification. The line-wise evaluation assesses the match between predicted and actual locations at the line level. 3) Code-wise evaluation based on top-k line predictions. It evaluates whether the bug in a code sample is correctly identified. It is based on the correctness of the top-1 and top-5 line predictions within each code sample. All evaluations utilize binary classification metrics such as  Precision, Recall, F1 score, and the area-under-curve (AUC) of the ROC curve~\cite{bradley1997use}.

The precision measures the ratio of correctly identified bug locations among all predicted locations, and recall evaluates the proportion of correctly identified bug locations against actual bug locations. It's worth noting the overall number of buggy tokens/lines is far less than non-buggy tokens/lines. In other words, the overall dataset is skewed by class bias. Therefore, it is important to report both the precision and recall of each model for a fair evaluation.

\textbf{Bug Correction Generation Metrics.}
To evaluate the performance of LLMs in generating bug corrections, we evaluate whether LLM-generated code snippets align with the desired functionality. Since LLM decoders are generative models, the outputs they produce are in a flexible format, with some contents that are not code but describe the underlying structure and semantics of the code. Thus, evaluation metrics that are too strict may underestimate the real performance of LLMs. To better evaluate the quality of the LLMs' ability on bug correction generation, we employ a `strict substring match' evaluation metric. This metric compares the processed\footnote{The string will undergo processing to eliminate spaces and special characters.} output from the LLMs with the correct label code snippet, trying to capture a strict match\footnote{The label (correct snippet) will be used to strictly match with output of the LLMs within a limited length. Here we limit the LLMs' generation length to a maximum 3 times of the length of the snippet label.}.

\subsection{Baseline Selection}
\label{sec:baseline-select}

We use GPT-3.5 and GPT-4 \cite{achiam2023gpt}, the most widely used commercial LLM solutions, as commercial baseline models for comparison. For open-source LLMs, we experimented on various models, including Gemma-7B, Mistral-7B \cite{jiang2023mistral}, DeepSeek \cite{guo2024deepseek}, and Code-Llama \cite{roziere2023code}. 
Additionally, to evaluate the impact of bug correction loss on the bug identification performance, we evaluate the performance of HLSEncoder (the Encoder fine-tuned with only bug identification loss) as the ablation study.

\section{Experimental Result} \label{sec:expr}

\subsection{Bug Identification Performance}\label{sec:bug-identification-performance}
Table \ref{tab:bug-identification-overall} presents the comparison of bug identification performance at token-level, line-level, and code-level, respectively. Results show that HLSDebugger significantly outperforms both the open-source LLM (Gemma-7B) and the commercial GPTs \cite{achiam2023gpt} across all three granularities. 
Even with the significant improvement of the supervised dataset on improving Gemma-7B, HLSDebugger still achieves superior performance than Gemma-7B-SFT, demonstrating the effectiveness of our encoder-based bug identification solution and debugging training scheme.

\begin{table}[!b]
\centering
\resizebox{.46\textwidth}{!} {
\begin{tabular}{|c|c|c|}
    \hline
    \multirow{2}{*}{Model} & Bug Correction  & \multirow{ 2}{*}{Model Type}   \\
          &   Accuracy   &     \\  
    \hline
    \hline
    GPT-3.5 \cite{achiam2023gpt}& 8.5\% & \multirow{2}{*}{Commercial Model} \\
    GPT-4 \cite{achiam2023gpt}& 10.5\% & \\
    \hline
    \multirow{2}{*}{Gemma-7B~\cite{team2024gemma}} & \multirow{2}{*}{7.5\%} & Open-sourced model\\
    & & Not fine-tuned \\
 
    \hline
    Gemma-7B UFT & 6.3\% & Fine-tuned \\
  
    Gemma-7B SFT  & 15.1\% & on our dataset \\
    \hline
    \multirow{2}{*}{HLSEncoder + GPT-4} & \multirow{2}{*}{18.9\%} & Commercial model \\
    & &  + our encoder  \\
    \hline
    HLSEncoder  & \multirow{2}{*}{23.7\%} & SFT Gemma 7B\\
    + Gemma-7B SFT & & + our encoder \\
    \hline
    \textbf{HLSDebugger} & \textbf{37.6\%} & Our solution\\
    \hline   
\end{tabular}
}
\caption{Bug correction accuracy comparison. HLSDebugger outperforms GPT-4 by more than $3\times$.}
\label{tab:acc-no-loc}
\end{table}

Specifically, for the token prediction (`Token-wise accuracy' in Table~\ref{tab:bug-identification-overall}), HLSDebugger shows a significant improvement over GPT-4~\cite{achiam2023gpt}. Its precision of 54.3\% is more than four times higher, and the recall of 83.5\% is more than double that of GPT-4~\cite{achiam2023gpt}. The high recall indicates that the HLSDebugger can capture most of the bugs in the code. The HLSEncoder achieves a recall rate that is nearly comparable to that of the HLSDebugger (79.8\% versus 83.5\%), suggesting that HLSEncoder has a similar capability to HLSDebugger in identifying bugs. However, HLSEncoder has a lower precision (46.1\% compared to 54.3\%), indicating that HLSDebugger is more effective at accurately understanding detailed information.

For line-wise prediction (`Line-wise accuracy' in Table~\ref{tab:bug-identification-overall}), experiments show a similar trend as in token-wise evaluation. In practice, locating the specific buggy line of code is sufficient for identifying the problem. HLSDebugger still achieves over two times higher recall and precision than GPT-4~\cite{achiam2023gpt}, indicating greater robustness and consistency. At the higher level (line-wise) identification, the HLSEncoder shows lower performance than HLSDebugger, but significantly higher than Gemma-7B SFT, which aligns with the results on the `token-wise accuracy'.

For code-wise evaluation based on top-k code line predictions (`Code-wise accuracy' in Table~\ref{tab:bug-identification-overall}), the HLSDebugger notably excels, particularly with the top-1 code line. As defined in Section \ref{sec:identification-metric}, top-1 accuracy measures whether the highest probability code line prediction is correct,  while top-5  accuracy requires only one of the top-5 line predictions to be correct. Consequently, top-5 accuracy is expected to be substantially higher than top-1. Notably, the accuracy for top-1 and top-5 lines of HLSDebugger reaches 79.5\% and 87.6\% for code-level evaluation, indicating that our model can successfully identify bugs in most buggy code samples with 1 to 5 trials.

\begin{figure}[!t]
\subfigure[Bug identification precision of different LLMs on all bug types.]
{
\centering
\includegraphics[width=0.46\textwidth]{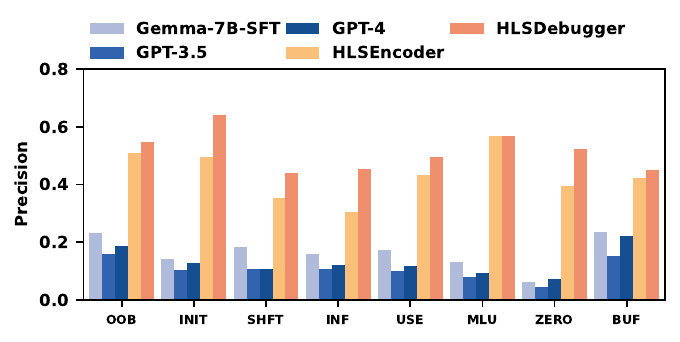}\label{fig:bug-identification-precision-four-model}\label{fig:token-pred-precision}
}
\subfigure[Bug identification recall of different LLMs on all bug types.]
{
\centering
\includegraphics[width=0.46\textwidth]{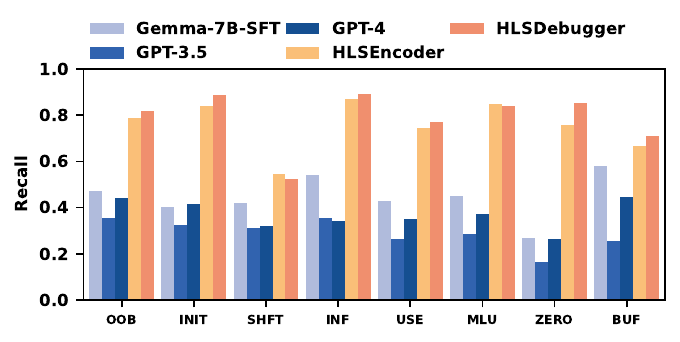}\label{fig:bug-identification-recall-four-model}\label{fig:token-pred-recall}
}
\caption{Bug identification performance of different LLM solutions on all bug types, including HLSEncoder for ablation study.} 
\label{fig:token-pred-bug-type}
\vspace{-.08in}
\end{figure}

\subsection{Bug Correction Performance}\label{sec:bug-correction-performance}

In addition to purely identifying bugs, an even more challenging task is to correct the bugs in a given code sample $C_\text{buggy}$. No information about bug locations is provided. Table~\ref{tab:acc-no-loc} shows the performance of LLM solutions in bug correction tasks. Despite Gemma-7B-UFT's enhanced bug identification (Section~\ref{sec:bug-identification-performance}), fine-tuning with an unsupervised dataset reduces its performance to 6.3\% on bug correction, compared to 7.5\% with Gemma-7B. HLSDebugger accurately corrected bugs in 37.6\% of given buggy code samples, more than 3 times better than commercial tool GPT-4's 10.5\%.
The supervised fine-tuning improves the correction accuracy of Gemma-7B from 3.5\% to 15.1\% (Gemma-7B SFT).
The HLSEncoder further improves the performance of Gemma-7B SFT by 7.6\%, ranking 2$^\text{nd}$ with 23.7\%.
Additionally, the HLSEncoder improves the performance of GPT-4~\cite{achiam2023gpt} by 8.4\%, ranking 3rd with 18.9\%.
The above results demonstrate the effectiveness of both our proposed dataset and HLSEncoder.
The gap between HLSDebugger and HLSEncoder+Gemma-7B-SFT (37.6\% v.s. 23.7\%) implies the advantage of the encoder-decoder structure that combines bug identification with correct tasks.

Overall, we admit that all existing LLMs, including HLSDebugger with the best 37\% accuracy, are not yet ready for practical debugging applications. But the obvious 3$\times$ superior performance of HLSDebugger over GPT-4 implies room for improvement for specific debugging tasks with customized models. With a larger and higher-quality training dataset, as well as a larger-scale and more LLM structure, we expect a better performance of LLM-aided HLS debugging.  \looseness=-1

\section{Discussion}   \label{sec:disc}

In this section, we present more detailed results of HLSDebugger on each individual bug type. We also report detailed results of GTP-4 \cite{achiam2023gpt} and Gemma-7B-SFT, the two strongest baselines.


We further provide detailed results on each of the 8 types of bugs. The detailed results cover two parts:  1) Token-wise bug identification, 2) Bug correction accuracy.

\textbf{Token-wise bug-identification.} Figure~\ref{fig:token-pred-bug-type} shows detailed results on identification (a) precision and (b) recall for each individual bug type. The accuracy of HLSDebugger significantly exceeds GPT-4~\cite{achiam2023gpt} and open-source models across all bug types.
This superior performance is consistent with the overall results in Table~\ref{tab:bug-identification-overall}.

\begin{figure}[!t]
    \centering
    \includegraphics[width=0.5\textwidth]{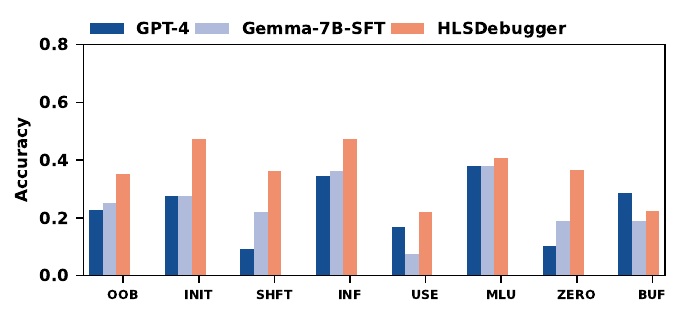}
    \caption{Bug correction accuracy of three models on different bug types. The blue bars correspond to normal bugs, while the orange bars correspond to pragma-related bug types.}    \label{fig:bug-correction-three}
\end{figure}

\textbf{Bug Correction Accuracy.}
Figure~\ref{fig:bug-correction-three} shows a detailed bar plot of the bug correction accuracy of various LLM solutions and our HLSDebugger on each specific bug type.
%
%
We can observe that all three models show similar trends across different bug types. 
Gemma-7B-SFT slightly outperforms GPT-4~\cite{achiam2023gpt} in bug corrections. 
HLSDebugger demonstrates superior performance to both GPT-4~\cite{achiam2023gpt} and Gemma-7B-SFT across the majority of the 8 bug types.

\begin{table}[!t]
    \centering
    \resizebox{.4\textwidth}{!}
    {
    \begin{tabular}{|c|c|}
          \hline
          
        Model & Bug Correction Accuracy  \\
        \hline
        \hline
        GPT-3.5 \cite{achiam2023gpt}& 36.0\%  \\
        GPT-4 \cite{achiam2023gpt}&  43.6\% \\
        Gemma-7B-UFT \cite{team2024gemma}& 14.7\%  \\ 
        Gemma-7B-SFT \cite{team2024gemma}& 45.4\%  \\
        \hline
        \textbf{HLSDebugger} & \textbf{49.1\%}  \\
        \hline
    \end{tabular}
    }
    \caption{ Bug correction accuracy given ground-truth bug location.}
    \label{tab:acc-with-location}
    \vspace{-.2in}
\end{table}

\textbf{Bug Correction Accuracy Given Location.}
In addition to bug identification, another contribution of HLSDebugger is bug correction. We add a new experiment in Table~\ref{tab:acc-with-location} to evaluate the bug correction accuracy, assuming the correct bug location is given (i.e., assuming 100\% identification accuracy). Given the ground-truth bug identifications, all LLMs' performance has been improved. HLSDebugger still achieves the highest performance 49.1\% in this sole bug correction task.



\section{Conclusion}\label{sec:concl}

In conclusion, we target the problem of debugging HLS logic bugs, 
addressing three challenges: 
1) scarcity of circuit data, 2) challenge of logic bugs, 3) multi-tasking in HLS debugging.
We propose the HLSDebugger, a customized LLM-based method for HLS logic debugging tasks. HLSDebugger first generates a large-scale labeled dataset, then adopts a customized structure, and also proposes a debugging training scheme.
Our solution achieved higher performance on both bug identification and correction generation compared with commercial LLM solutions. 



\section{Acknowledgement}
This research was supported by Hong Kong Research Grants Council (RGC) CRF-YCRG Grant C6003-24Y, GRF 16200724, and ACCESS – AI Chip Center for Emerging Smart Systems, sponsored by the InnoHK initiative of the Innovation and Technology Commission of the Hong Kong Special Administrative Region Government.



\ifCLASSOPTIONcaptionsoff
  \newpage
\fi

\bibliographystyle{IEEEtran}
\bibliography{ref}





\end{document}